\documentclass[aps,prd,a4paper,preprintnumbers,twocolumn,floatfix,showkeys,showpacs,nofootinbib]{revtex4}

\usepackage{psfrag}
\usepackage{amssymb}
\usepackage{subfigure}
\usepackage{color}
\usepackage{mathrsfs}
\usepackage{graphicx}

\def\be{\begin{equation}}
\def\ee{\end{equation}}
\def\bea{\begin{eqnarray}}
\def\eea{\end{eqnarray}}

\begin{document}


\title{Are quantization rules for horizon areas universal?}



\author{Valerio Faraoni}
\email[]{vfaraoni@ubishops.ca}
\affiliation{Physics Department and {\em STAR} Research Cluster, 
Bishop's University, 2600 College Street, 
Sherbrooke, Qu\'ebec, Canada J1M~1Z7}

\author{Andres F. Zambrano Moreno}
\email[]{azambrano07@UBishops.ca}
\affiliation{Physics Department,  
Bishop's University, 2600 College Street, 
Sherbrooke, Qu\'ebec, Canada J1M~1Z7}



\begin{abstract} 
Doubts have been expressed on the universality of 
holographic/string-inspired quantization rules for the 
horizon areas of stationary black holes, or the products 
of their radii, already in  4-dimensional general  
relativity.  Realistic black holes are not stationary but 
time-dependent. We produce three examples of 4D 
general-relativistic spacetimes containing dynamical black 
holes for at least part of the time, and we show  that the 
quantization rules (even counting virtual horizons) cannot 
hold, except possibly  at isolated instants of time, and do 
not seem to be universal. 
\end{abstract}


\maketitle


\section{\label{section1}Introduction}

Recently, there has been some excitement in the research 
community working on the holographic principle and  
stringy/supergravity black holes following the observation 
that the products of Killing horizon areas for certain 
multi-horizon black holes are independent of the black hole 
mass and depend only on the quantized charges (supergravity 
and extra-dimensional black holes with angular momentum and 
electric and magnetic charges were considered) 
\cite{Larsen97, CveticLarsen97, CveticGibbonsPope,  
AnsorgHennigPRL, 
AnsorgHennigCQG, CastroRodriguez, expression, Meessenetal, 
Castronew}. 
Older results on black holes far from 
extremality \cite{Curir79, Larsen97, CveticLarsen97}  induce one 
to take into account both outer 
and inner black hole horizons when studying the 
quantization of  black hole entropies and horizon areas. 
Expressions for  products of the horizon areas of black 
holes in four and higher dimension have been hypotesized or  
suggested \cite{CveticGibbonsPope, AnsorgHennigPRL, 
AnsorgHennigCQG, 
CastroRodriguez} and  then questioned in more recent 
work 
\cite{Castronew}. 

This literature is inspired by the 
holographic principle and string theories (although the 
results are not, strictly speaking, derived from string theories), 
and it stems from the underlying idea that quantized 
products of 
areas depending on combinations of integers must carry the 
signature of some specific microphysics. This feature would 
not be too surprising if the area $A$ of an horizon is 
related to its entropy $S$ through the famous 
Bekenstein-Hawking formula $S=A/4$ (in units in which 
$c=\hbar=1$) and corresponds to a statistical mechanics 
based on microscopic models counting microstates determined 
by quantum gravity (see, {\em e.g.}, ref.~\cite{Sen08}). 
When there are outer ($+$) and inner 
($-$) horizons, the quantization rules recurrent in the 
literature are 
\be\label{expression1} 
A_{\pm} = 8\pi 
l_{pl}^2 \left( \sqrt{N_1} \pm \sqrt{N_2} \, \right) 
\,, \;\;\;\;\;\;\;\; N_1, N_2 \in \mathbb{N} \,, 
\ee 
or 
\be\label{expression2} 
A_{+} A_{-} =\left( 8\pi 
l_{pl}^2\right)^2 N \,, \;\;\;\;\;\;\;\; N \in \mathbb{N} 
\,, 
\ee 
where $l_{pl}$ is the Planck length 
\cite{Curir79, Larsen97, CveticLarsen97}. $N_{1,2}$ are integers for 
supersymmetric extremal black holes but are related to the 
numbers of branes, antibranes, and strings in less simple 
situations \cite{HorowitzMaldacenaStrominger}. A weaker 
rule states that the product of horizon areas is 
independent of the black hole mass and depends only on the 
quantized charges.  Rules of the type~(\ref{expression2}) 
are found for Einstein-Maxwell black holes in 5 and 6 
dimensions, asymptotically flat \cite{Curir79, Larsen97, 
AnsorgHennigCQG, AnsorgHennigPRL, Meessenetal, 
CastroRodriguez}, 
or 
asymptotically 
de Sitter or anti-de~Sitter, and also for black holes in 
$D=3$  and $D\geq 6$ dimensions \cite{Castronew} and it 
seems to apply also to black rings and black strings in 
higher dimension \cite{CastroRodriguez} (asymptotically de 
Sitter and anti-de Sitter black holes in general relativity 
and other theories of gravity, in various dimensions, are 
discussed in \cite{Castronew}).

A word of caution against the  
temptation of regarding these rules as universal for all 
types of black holes endowed with multiple horizons  
\cite{CveticGibbonsPope} has 
been voiced  by Visser \cite{Visser0, Visser}. Visser 
considered black holes in 
4-dimensional general relativity and found that, in these 
situations, products of areas do not give mass-independent 
quantities, nor are they related in a simple way to 
integers. Rather, it is quadratic combinations of the 
various horizon radii (with the dimensions of an area, 
which can be referred to as ``generalized areas'') which 
generate mass-independent quantities and are, presumably, 
the best candidates to be quantized \cite{Visser0, Visser}, 
although no evidence has been presented thus far that these 
generalized areas have any special physical  significance. 
Moreover, it is essential to include in these algebraic 
combinations also cosmological and virtual horizons in 
addition to the black hole horizons \cite{Visser}. Virtual 
horizons correspond to negative or imaginary roots of the equation 
locating the horizons (which, in non-asymptotically flat 
solutions of the Einstein equations, provides also 
cosmological horizons). The quantization rules break down 
also for general Myers-Perry black holes in dimension $D 
\geq 6$ and for Kerr-anti-de Sitter black holes with $D 
\geq 4$ unless the virtual horizons are included in the 
picture \cite{ChenLiu}.

In this note we point out a fact which induces even more caution 
in discussing the products of horizon areas.
The horizons considered in the literature are Killing (and event) 
horizons. Realistic black holes are not 
stationary if 
nothing else because they emit Hawking radiation and the 
backreaction due to this effect changes their masses which
become time-dependent, together with their horizon radii and areas. For astrophysical 
black holes the effect is completely negligible but the same 
cannot be said for quantum black holes. Therefore, a timelike 
Killing 
vector will not 
be present and in realistic situations  one should consider 
not Killing and 
event horizons, but other kinds of horizons. Dynamical horizons have received much attention
in quantum gravity \cite{dynamicalhorizons}; at present it 
seems that 
apparent horizons (``AH''s, see \cite{AHreviews} for reviews) are the best and most versatile candidates 
for the notion of time-dependent ``horizon'' 
and it is claimed that thermodynamical laws can be associated with AHs
\cite{AHthermodynamics}. In any case, AHs are used as proxies for event horizons in 
studies of gravitational  collapse in  numerical relativity 
\cite{proxies}. AHs coincide with  event horizons in 
stationary situations but, in dynamical situations, they 
are  spacelike  or even timelike. In the following 
we  consider dynamical situations and we focus on AHs.

\section{\label{section2}Toy models for dynamical black holes}

Here we consider three toy models of {\em dynamical} black 
holes, which are implemented by embedding  
them in a Friedmann-Lema\^{i}tre-Robertson-Walker (FLRW) 
cosmological ``background'' (we use quotation marks 
because, due to the non-linearity of the Einstein 
equations, one cannot split the metric into a background 
and a deviation from it in a covariant way). In the first 
model, a McVittie spacetime, there are a black hole, a cosmological, and a virtual horizon. In 
the second model, a generalized McVittie  
solution of the Einstein equations, the ``McVittie 
no-accretion condition'' is relaxed to allow accretion of 
energy and then we have either two real  
horizons (a black hole and a cosmological horizon) or two 
virtual horizons. The third model consists of an 
electrically charged (but non-accreting) generalization of 
the McVittie spacetime. In this case there is  a 
charge to quantize, but the behaviour of the horizons is the same as 
in the uncharged case.  Our main point is that, in 
dynamical situations, even if combinations of AH radii which are 
mass-independent exist, they depend continuously on time 
and cannot be expressed as combinations of integers.

\subsection{\label{subsec:McV}The McVittie spacetime}

\begin{figure}[t]
\includegraphics[scale=0.9]{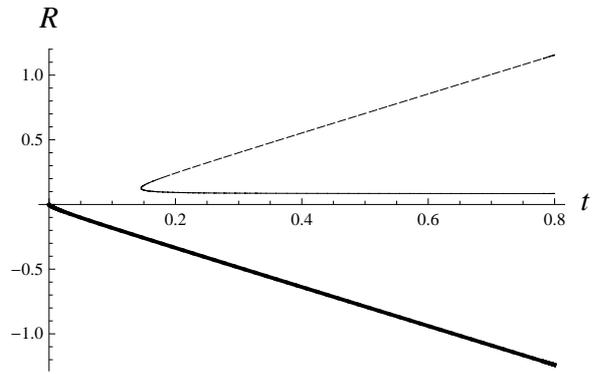}
\caption{\label{figure1}The proper radii of the AHs of a 
dust-dominated McVittie metric versus time. The negative 
radius 
represents the virtual horizon. At a  critical time a 
cosmological AH (dashed curve) appears together with a 
black hole AH (solid curve), the  former expanding and the 
latter shrinking.} 
\end{figure}

The McVittie metric  \cite{McVittie} describes a black hole 
embedded in a FLRW universe, which is 
a truly dynamical spacetime.\footnote{The special case 
of 
a de Sitter ``background'' admits a timelike Killing 
vector and is locally static in the region between the 
black hole and the de Sitter cosmological horizons.}  
Limiting ourselves, for 
simplicity, to a spatially flat FLRW ``background'', 
the line element can thus be 
written in the form \cite{Roshina} 
\begin{eqnarray}
ds^2 &=& -\left[ 1-\frac{2m}{R}-H^2(t)R^2 \right] dt^2 
+\frac{dR^2 }{1-\frac{2m}{R}} \nonumber\\
&&\nonumber\\
&\, & -\frac{2H(t) R}{\sqrt{1-\frac{2m}{R}}} \, dtdR 
+R^2d\Omega_{(2)}^2 \,, \label{McVittie} 
\end{eqnarray}
where $m$ is a constant related to the mass of the  
central inhomogeneity, 
$d\Omega_{(2)}^2=d\theta^2+\sin^2 \theta \, d\varphi^2$ is 
the metric on the unit 2-sphere, $R$ is the areal radius, 
$H(t) \equiv \dot{a}(t)/a(t)$ is the Hubble parameter, $a(t)$ is the 
scale factor of the FLRW ``background'', and an overdot 
denotes differentiation with respect to the 
comoving time $t$.  The locally static Schwarzschild-de Sitter-Kottler spacetime corresponds to 
$a(t)=\exp(\sqrt{\Lambda/3} \, t)$ and $H=\sqrt{\Lambda/3}$  
(where $\Lambda>0$ is the
 cosmological constant) and is a special case of the 
McVittie metric which can be obtained using a simple transformation of the time coordinate 
 \cite{arakida}.  Assuming a perfect fluid stress energy 
tensor, the Einstein equations provide 
the energy density $\rho(t)$ and pressure $P\left(t, 
R\right)$ of the ``background'' fluid. Again 
for simplicity, let us restrict ourselves to a cosmic fluid which reduces 
to dust (equation of state 
parameter $w \equiv P/\rho=0$) at spatial infinity, then 
\be 
\rho(t)=\frac{3}{8\pi} \, H^2(t) 
\,, 
\ee 
\be
 P\left(t,R \right)=\rho(t) \left( 
\frac{1}{\sqrt{1-\frac{2m}{R}}} -1 \right) \label{pressure} \,.
\ee
The inverse metric is
\be
\left( g^{\mu\nu} \right)=\left(
\begin{array}{cccc}
  \frac{-1}{1-2m/R} &  \frac{-HR}{\sqrt{1-2m/R}} & 0 & 
0 \\
&&& \\
\frac{-HR}{ \sqrt{1-2m/R} } & \left(1-\frac{2m}{R} 
-H^2R^2  \right)   & 0 & 0 \\
&&&\\
0 & 0 & \frac{1}{R^2}  & 0 \\
&&&\\
0 & 0& 0 & \frac{1}{R^2 \sin^2 \theta} 
\end{array} \right)\,. \label{inverseMcVittie} 
\ee
For any spherically symmetric metric written in terms of the areal radius   $R$, 
the AHs are located by solving the equation $\nabla^cR\nabla_c R=0$ or $g^{RR}=0$ 
\cite{NielsenVisser}.   
For the  Schwarzschild-de Sitter-Kottler spacetime, which 
is a special case of McVittie, this equation coincides 
with the horizon condition reported in \cite{Visser} but, 
in the general case,  the  Hubble  parameter is 
time-dependent instead of constant. This cubic equation 
\be
R^3 -\frac{R}{H^2(t)} + \frac{2m}{H^2(t)}=0 
\ee
has three solutions which, under conditions specified 
below, correspond to a
 time-dependent black hole AH with (proper) radius $R_{BH}(t)$, a cosmological AH with radius 
$R_C(t)$, and a virtual AH with negative radius $R_V(t)$. 
The three roots are 
\begin{eqnarray}
R_{BH}&=&\frac{2H^{-1}}{\sqrt{3}}\sin\psi 
\,,\label{root1}\\
&&\nonumber\\
R_C&=& -R_V= H^{-1} \left( 
\cos\psi -\frac{1}{\sqrt{3}}\sin\psi \right) 
\,,\label{root2}
\end{eqnarray}
with $\psi (t)$  given by  $\sin (3\psi )=3\sqrt{3} \, 
mH(t)$. Here $m$ and $H$ 
are  both necessarily positive (we only consider expanding 
universes) and $R_V$ defines the negative root. As 
discussed in \cite{AndresRoshina}, the 
condition for the black hole and cosmological AHs to exist is $0<\sin 
( 3\psi ) <1$, which corresponds to  
$mH(t)<1/(3\sqrt{3})$ (and $mH(t)>0$, which is always 
satisfied).  Unlike the Schwarzschild-de Sitter-Kottler case  where the Hubble parameter 
is a constant, this inequality will only be satisfied at 
certain times during the cosmological expansion and will be violated at  
other times. The threshold between these two regimes is the 
time at which $ m H(t_* )=1/(3\sqrt{3})$ (for a 
dust-dominated ``background'' with $H(t)=2/(3t)$, 
this critical time is $t_* =2\sqrt{3} \, m$).  At early times 
 $t<t_*$ it is  $m>\frac{1}{3\sqrt{3} 
\,H(t)}$ and both $R_{BH}(t)$ and $R_C(t)$ are complex and therefore 
unphysical. In this case all the AHs are virtual. 
At the critical time $t=t_*$ it is 
$m=\frac{1}{3\sqrt{3}\,H(t)}$ 
and the AHs with radii $R_{BH}(t_*)$ and $R_C(t_*)$ 
coincide at a real,  physical location. 
There are then a single real AH at $\frac{1}{\sqrt{3}\,H(t)}$ and one virtual AH.
 At ``late'' times  $t>t_*$ it is  $m<\frac{1}{3\sqrt{3}\,H(t)}$, 
and  both $R_{BH}(t)$ and $R_C(t)$ are real and, therefore,  
physical~---~there are 
two real and one virtual AHs. The dynamics of the black hole and cosmological AH radii
 as functions of comoving time are pictured in  
fig.~\ref{figure1}. 

The phenomenology of AHs appearing and annihilating in 
pairs appears to be rather common  
for black holes embedded in cosmological ``backgrounds'', 
in both general relativity and alternative
theories of gravity  \cite{HMN, myClifton, CMB}. The  
physical reason why a pair of AHs suddenly 
appears
in the McVittie spacetime (\ref{McVittie}) is discussed in  
\cite{AndresRoshina}. The same phenomenology 
of fig.~\ref{figure1} is found for generalized McVittie 
metrics \cite{Gaoetal} and in 
Lema\^itre-Tolman-Bondi spacetimes  (\cite{GaoLTB}, see 
also \cite{othersplittingAHs}) 
describing black holes embedded in 
(spatially flat) FLRW universes.\footnote{In the first case both 
decelerating and 
accelerating FLRW ``background'' universes are considered 
while in the second case, by necessity, 
only a dust-dominated ``background'' is considered.}

The Misner-Sharp-Hernandez mass $M_{MSH}$  
\cite{MisnerSharpHernandez} of a sphere of areal radius  
$R$ (which is defined for spherically symmetric 
spacetimes) is \cite{AndresRoshina} 
\be 
M_{MSH}= m + \frac{4\pi G}{3} \, \rho \, R^3  
\label{MisnerSharpmass}
\ee
and coincides with the Hawking-Hayward quasi-local mass 
\cite{HawkingHayward} in spherical symmetry. 
It is interpreted as the contribution of the black hole mass $m$ (which 
is constant because of the ``McVittie condition'' $G^0_1=0$, which implies $T^0_1=0$ for 
the stress-energy tensor of the cosmic fluid and forbids accretion of the latter 
onto the black hole) and a contribution due to the energy of the cosmic fluid inside 
the sphere. Searching for generalized areas which are independent of the black hole mass, 
Visser's discussion for the Schwarzschild-de Sitter-Kottler black hole can be repeated 
almost without changes. Including the virtual horizon in the count,  it is straightforward to 
see that the quantities
\be\label{massindependent1}
R_V\left( R_{BH} + R_C \right)+R_{BH}R_C = -\frac{1}{H^2(t)} 
\ee
and
\be\label{massindependent2}
\left( R_{BH}+R_{C} \right)^2-R_{BH}R_C =\frac{1}{H^2(t)}
\ee
are independent of the black hole mass $m$. This situation can be regarded as a special 
case of Visser's discussion  
 \cite{Visser} computing  mass-independent combinations of 
AH radii whenever the 
Misner-Sharp-Hernandez mass is a Laurent polynomial of the areal radius $R$. This is clearly 
the case of the McVittie metric, see eq.~(\ref{MisnerSharpmass}). In the present case, the 
physical mass contained in a sphere is actually given by the Misner-Sharp-Hernandez notion, but 
the cosmic fluid here serves the only purpose of generating 
a cosmological ``background'' to make the 
central black hole dynamical and it seems that the relevant mass to consider when  
mass-independent quantities such as (\ref{massindependent1}) and (\ref{massindependent2}) 
are searched for is the black hole contribution $m$, not the total $M_{MSH}$. In any case, the 
AH radii identify different spheres and correspond to different Misner-Sharp-Hernandez masses 
$M_{MSH}^{(i)}=2R^{(i)}_{AH} $ (from eq.~(\ref{MisnerSharpmass})). Here we stick to $m$.

Following \cite{Visser}, we have included the virtual 
horizon to obtain the mass-independent 
quantity~(\ref{massindependent1}). Now, when the AH radii 
change with time, the combinations (\ref{massindependent1}) 
and (\ref{massindependent2}) are not constant but depend on 
time: therefore, if they are expressed by combinations of 
integers at an initial time, they will not be combinations 
of integers immediately afterward. They could only be a 
combination of integers at times forming a set of zero 
measure in any time interval.

\subsection{\label{subsec:SD}The generalized (accreting) 
McVittie spacetime}
 
\begin{figure}[t]
\includegraphics[scale=0.7]{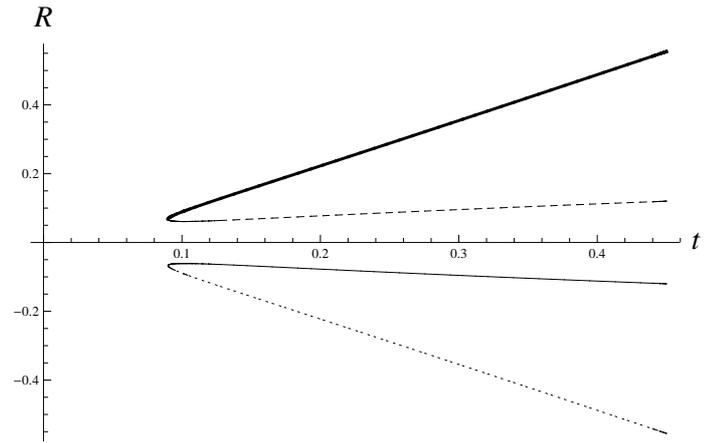}
\caption{\label{SDfigure}The radii of the AHs of the 
generalized (accreting) McVittie spacetime versus time.   
There are always either two  virtual AHs (with negative 
radii) or two real AHs with $R\geq 0$ (a cosmological AH, 
thick solid curve, and a black hole AH, dashed curve, both 
appearing at a critical time).} 
\end{figure}

The McVittie solution of the Einstein equations can be 
generalized to allow for the possibility of radial energy 
flow onto the central inhomogeneity 
\cite{FaraoniJacques, Gaoetal}. Among the class of 
spherically symmetric 
solutions of the Einstein equations thus obtained there is 
a late-time attractor for  expanding ``background'' 
universes which is given by the line element 
\cite{generalMcV} 
\begin{eqnarray}
ds^2 &=&  - \frac{
\left( 1-\frac{m}{2r} \right)^2}{ 
\left( 1+\frac{m}{2r} \right)^2} dt^2 \nonumber\\
&&\nonumber\\
&\, & +a^2(t) \left( 1+\frac{m}{2r}\right)^4 \left( dr^2 
 +r^2 d\Omega_{(2)}^2 \right)  
\end{eqnarray}
in isotropic coordinates, where $a(t)$  is the scale 
factor of the ``background'' FLRW universe and $m$ is a  
constant. In terms of the areal radius $R(t,r)=a(t)r\left( 
1+\frac{m}{2r}\right)^2$, the AHs of this solution of the 
Einstein equations corresponding to an expanding FLRW 
``background'' universe are \cite{generalMcV} 
\begin{eqnarray}
R_{BH} & = & \frac{1-\sqrt{1-8MH } }{2H} \,,\label{Root1}\\
&&\nonumber\\
R_{CH} & = &\frac{1+\sqrt{1-8MH } }{ 2H}  \,, \label{Root2}
\end{eqnarray}
where $M(t)=m \, a(t)$ and $H\equiv \dot{a}/a$ is the Hubble 
parameter of the  ``background'' FLRW universe \cite{generalMcV}. 
The time evolution of the radii of the apparent horizons is 
reported in fig.~\ref{SDfigure} for a dust-dominated FLRW 
universe with scale factor $a(t)=a_0 t^{2/3}$.   
 These AHs are real when $8M H\leq1$ and virtual when $8M H 
> 1$ (in which case we label them $R_{V1}$ and $R_{V_2}$), 
which happens before a critical time $t_*$.

The products of the horizon radii prescribed in  
\cite{Visser} reduce to  
\be 
R_{CH} R_{BH} = R_{V_1} R_{V_2}= \frac{2M}{ H(t) }=3m a_0 
t^{5/3} \,, \label{productSD} 
\ee 
which is time-dependent, for both cases in which the AHs 
are real or virtual. Following the same reasoning as in the 
previous section, we conclude that it cannot be expressed 
as a combination of integers. The 
Misner-Sharp-Hernandez/Hawking-Hayward mass of the  black 
hole (when the latter exists) is \cite{generalMcV}
\be
M_{MSH}=\frac{ R_{BH} }{2}=\frac{1-\sqrt{1-8MH}}{4H} \,;
\ee
it cannot be split in any simple way into a contribution 
due to the central inhomogeneity and one due to the cosmic 
fluid inside the sphere of radius $R_{BH}$. It is time-dependent 
due to the radial energy flow onto the black hole.  The product 
(\ref{productSD}) can be rewritten as
\begin{eqnarray}
R_{CH} R_{BH} &=&  R_{V_1} R_{V_2} = 
2M_{MSH} \left( \frac{1+\sqrt{1-8MH}}{2H} \right) 
\nonumber\\
&& \nonumber\\
&= & 2M_{MSH} R_{CH}
\end{eqnarray}
and depends on the physical black hole mass.

\subsection{\label{subsec:McVcharged}The electrically 
charged McVittie spacetime}

Let us consider now the electrically charged, non-accreting,  
generalization of the McVittie spacetime. This situation is 
physically more interesting because there is actually a charge 
which could be quantized. The electrically charged McVittie 
spacetime is described by the spherically symmetric line 
element \cite{GaoZhang04}
\begin{eqnarray}
ds^2 &=& -\frac{\left( 1-\frac{m^2}{4a^2r^2}\right)^2  
+\frac{Q^2}{4a^2r^2}}{ \left[ \left( 
1+\frac{m}{2ar}\right)^2 -\frac{Q^2}{4a^2r^2}\right]^2  }  
\, dt^2  \\
&&\nonumber\\
&\,& + a^2(t) \left[
\left( 1+\frac{m}{2ar} \right)^2 -\frac{Q^2}{4a^2r^2} 
\right]^2 \left( dr^2 + r^2 d\Omega_{(2)}^2 
\right)\nonumber 
\end{eqnarray}
in isotropic coordinates, where $m=$~const.  is a mass 
parameter, $Q$ is the electric 
charge of the central inhomogeneity, and $a(t)$ is the  
scale factor of the ``background'' FLRW universe, which is 
chosen here to be spatially flat.

\begin{figure}[t]
\includegraphics[scale=0.35]{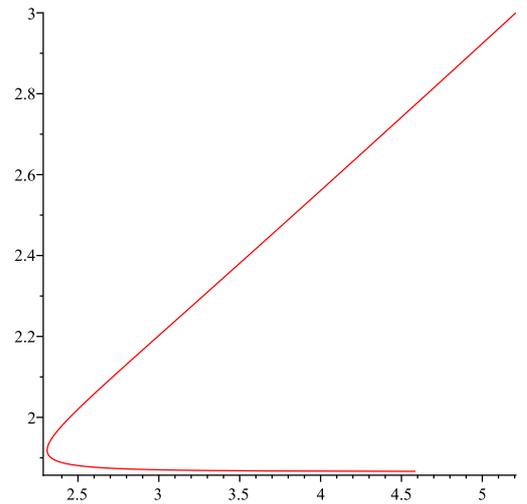}
\caption{\label{figchargedMcV}The proper radii versus time 
of the real AHs of an electrically charged McVittie spacetime with 
a spatially flat, dust-dominated, FLRW ``background''.} 
\end{figure}

The areal radius is clearly
\begin{eqnarray}
R\left(t,r \right) &=& a(t) r\left[ \left( 
1+\frac{m}{2ar}\right)^2 
-\frac{Q^2}{4a^2r^2} 
\right] \nonumber\\
&&\nonumber\\
&=& ar+m+\frac{m^2}{4ar}-\frac{Q^2}{4ar}\label{chargedArealRadius}
\end{eqnarray}
and the apparent horizons are located by the equation 
$\nabla^cR\nabla_cR=0$. After a straightforward 
calculation, this equation becomes 
\be\label{chargedAHr}
H^2  \left[ \left( 2ar+m \right)^2 - 
Q^2\right]^4 
-4 \left( 4a^2r^2 -m^2+Q^2 \right)=0 \,.
\ee 
This form is not particularly useful for locating the apparent horizons 
because it is expressed in terms of the radial  coordinate $r$; in order to
turn it into a more useful expression involving only the proper 
(areal) radius $R$ and the time $t$, one inverts eq.~(\ref{chargedArealRadius}) 
and obtains 
\be\label{2ar}
r=\frac{1}{2a}\left( R-m+ \sqrt{ R^2+Q^2-2mR } \right)
\ee
by choosing the positive sign of the square root. 
Using eq.~(\ref{2ar}),  eq.~(\ref{chargedAHr}) becomes 
\begin{eqnarray}
4H^2(t) R^4 \left( R-m+ \sqrt{ R^2+Q^2-2mR} \right)^4 
\nonumber\\
\nonumber\\
- \left( R-m+ \sqrt{ R^2+Q^2-2mR} \right)^2 + (m^2-Q^2)=0 \nonumber\\
 \label{chargedAHR}
\end{eqnarray}
in terms of $R$ and $t$. It is complicated to solve this trascendental equation or even to give explicit 
analytical criteria for the existence and number of its roots but it is clear that, when 
solutions exist, they depend on time and generalized areas and their products will also 
depend on time and will not be expressed as simple combinations of integers times a constant.
For illustration, eq.~(\ref{chargedAHR}) is  solved numerically for special values of the 
parameters using  
the scale factor $a(t)=a_0 t^{2/3}$ of a  dust-dominated 
FLRW  ``background'' and the radii of the real apparent horizons are plotted 
in fig.~\ref{figchargedMcV}.

The case of a charged McVittie spacetime with $|Q|=m$ can be treated explicitly as 
the 
relevant expressions simplify considerably in this case. The areal radius is simply
$ R=ar+m \geq m$ and the 
equation locating the apparent horizons becomes
\be
\frac{1}{16 H^2}= R^4 \left( R-m \right)^2
\ee
or
\be
S(R) \equiv R^2\left( R-m \right) = \frac{1}{4H}
\ee
in an expanding universe. The function $S(R)$ is a cubic with a local maximum of zero value at 
$R=0$ and a local minimum (of value $-4m^3/27$) at $R=2m/3$; in the physical range $ m \leq R < 
+\infty $ it is 
always increasing, starting from zero at $R=m$ and  going to infinity as $R\rightarrow +\infty$.  
Therefore, 
for any $t>0$ there is one and only one intersection between the graph of the function $S(R)$ and the 
horizontal line with ordinate value $1/4H= \frac{3t}{8}>0$ (where, as usual,
 we assume $a(t)=a_0t^{2/3}$ 
for a 
dust-dominated FLRW ``background''), {\em i.e.}, there is always one and only one 
apparent horizon with a radius which increases as the universe expands. (A detailed analysis of the 
apparent horizons of the charged McVittie spacetime, including the extremal case, will be reported 
elsewhere.)

\section{\label{section4}Conclusions}

The cosmological black holes reported here are just toy models for 
dynamical black hole horizons: the main point is that realistic black 
holes are time-dependent, not stationary. Therefore, far-reaching 
conclusions about the quantization of black hole horizon areas, or of 
quantities which are quadratic in the radii of Killing horizons 
(generalized areas), may be misleading and may not correspond to 
realistic, time-dependent, situations.  It is interesting to probe the 
conjecture about mass-independence and generalized area quantization using 
examples of time-varying black holes in 4-dimensional general relativity 
before approaching higher-dimensional black objects in  
supergravity or stringy objects. Exact 
solutions of the field equations of Einstein theory describing 
time-varying black holes are not easy to find and we resort to the more 
well known cosmological black holes to provide examples of time-dependent 
black holes---the cosmological ``background'' is not 
conceptually essential here. In 
general, AHs depend on the spacetime foliation \cite{foliationdependence} 
but in the presence of spherical symmetry, to which we have restricted 
ourselves, this does not appear to be a significant problem. 
For the McVittie metric (as well as for 
its special Schwarzschild-de Sitter-Kottler static case) there are 
generalized areas which are independent of the black hole mass. However, 
even if they can be expressed as $8\pi l_{pl}^2$ times a combination of 
integers at some initial time, this expression changes as time goes by. 
The corresponding quantities for the generalized accreting McVittie and 
the electrically charged McVittie black holes are mass- and 
time-dependent.

Variations on the theme can be contemplated. If only the 
black hole and the cosmological AHs are retained and 
considered as physical, their area will be zero at all 
times  $0<t<t_{*}$; zero is an integer, allright, but this 
interpretation entails an entropy 
suddenly jumping from zero (describing a naked singularity 
in a FLRW ``background'') to a 
value not reducible to a combination of integers and depending on the black hole mass. If 
the cosmological AH is excluded from the picture, then there remains only the black hole 
AH, the area of which is initially zero, then jumps to a positive value, and then 
decreases monotonically as time goes by (see 
fig.~\ref{figure1}).  More complicated black 
holes with multiple AHs will lend themselves to the consideration of more possible 
combinations of the AH radii, but probably the most 
sensible way to proceed is to include 
all AH radii, even virtual ones,  
when searching for quantizable, mass-independent quantitites, as done in \cite{Visser}. When realistic 
time-dependent horizons are considered, however, the connection between products of areas 
and combinations of integers becomes even more speculative and perhaps it would be better 
to put it on a firmer ground or finding out its limits of validity before assuming it as 
a postulate or a necessary accessory of the holographic principle. This conclusion 
reinforces that of Visser  \cite{Visser0, Visser} that the 
black holes of 4-dimensional 
general relativity do not seem to reconcile with the usual quantization rules 
(\ref{expression1}) and (\ref{expression2}) and casts 
serious doubts on the 
universality of these expressions.

\begin{acknowledgments}
VF is grateful to Matt Visser for a discussion. This work 
is supported by grants from Bishop's University and the 
Natural Sciences and Engineering Research Council of 
Canada.
\end{acknowledgments}



\end{document}